\documentstyle[12pt,times,cite,epsf]{article}
\def\Zslash{\hbox{$Z$\kern-0.5em\raise0.3ex\hbox{/}}}
\def\Wslash{\hbox{$W$\kern-0.5em\raise0.3ex\hbox{/}}}
\def\gt{c_L}
\def\cL{c_L}

\newcommand{\mysection}[1]{\setcounter{equation}{0}\section{#1}}

\newcommand{\beq}{\begin{equation}}
\newcommand{\eeq}{\end{equation}}
\newcommand{\bea}{\begin{eqnarray}}
\newcommand{\eea}{\end{eqnarray}}
\newcommand{\eq}[1]{eq.~(\ref{#1})}

\newcommand{\db}{\hspace{-0.2ex}\not\hspace{-0.7ex}D\hspace{0.1ex}}
\newcommand{\sla}[1]{\hspace{-0.1ex}\not\hspace{-0.5ex} #1\hspace{0.1ex}}
\newcommand{\bl}{b_L}
\newcommand{\bbl}{\overline{b}_L}

\newcommand{\msb}{$\overline{\mathrm{MS}}$ }
\newcommand{\zbb}{Z\rightarrow b\bar{b}}

\newcommand{\op}{{\cal O}}

\def\N{N_c}

\def\a{\alpha_s}
\def\mt{m_t}
\def\gf{G_F}
\def\G{{M_Z^3\sqrt 2 \gf\over 48 \pi}}
\def\4pi{(4 \pi)}
\def\dslash{\hbox{$\partial$\kern-0.5em\raise0.3ex\hbox{/}}}
\def\Dslash{\hbox{$D$\kern-0.5em\raise0.3ex\hbox{/}}}
\def\Aslash{\hbox{$A$\kern-0.5em\raise0.3ex\hbox{/}}}
\def\Gslash{\hbox{$G$\kern-0.5em\raise0.3ex\hbox{/}}}
\def\pslash{\hbox{$p\ $\kern-0.5em\raise0.3ex\hbox{/}}}
\def\m{\mu}
\renewcommand{\titlepage}{\clearpage%
\setcounter{footnote}{0}%
\thispagestyle{empty}\pagestyle{plain}\pagenumbering{arabic}%
\kern1mm
\vskip15mm\normalsize}
\newcommand{\docnum}[1]{\hbox to \hsize{\hskip123mm\hbox{#1}\hss}}
\renewcommand{\date}[1]{\hbox to \hsize{\hskip123mm\hbox{#1}\hss}}

\renewcommand{\title}[1]{\vskip1em\begin{center}\Large\bf#1\end{center}\vskip2.5em}
\renewcommand{\author}[1]{\vskip0.5em{\bf #1}\vskip0.5em}
\newcommand{\inst}[1]{\vskip0.3em{ #1}\vskip0.5em}
\renewcommand{\abstract}{\begin{center}{\bf Abstract}\end{center}\quotation}

\newcommand{\anotfoot}[2]{\vfill\noindent\underline{\hspace{6cm}}
\par\noindent #1) #2}
\newcommand{\anotfootnb}[2]{\par\noindent #1) #2}
\begin{document}
\begin{titlepage}
\docnum{CERN-TH/95-21}
\docnum{FTUV/95-9}
\docnum{hep-ph/9502307}
\vspace{2cm}
\title{An effective field theory approach \break
to the QCD corrections to the large-$m_t$ $Zb\bar b$ vertex}
\begin{center}
\author{Santiago Peris$^{*)}$}
\inst{TH Division, CERN, 1211 Gen\`eve 23, Switzerland}
\inst{and}
\author{Arcadi Santamaria$^{**)}$}
\inst{Departament de F\'{\i}sica Te\`orica, Universitat de Val\`encia,\\
E-46100 Burjassot, Val\`encia, Spain}
\end{center}

\vspace{2cm}

\begin{abstract}
Using effective field theory techniques we discuss the QCD corrections to
the large-$m_t$ contributions to the
process $Z\rightarrow b \bar b$. In particular we obtain the $\a$
correction to the non-universal $\log m_t$ contribution to the $Zb\bar b$
vertex.
\end{abstract}

\vspace{0.5cm}
\vfill\noindent
CERN-TH/95-21

\noindent
FTUV/95-9
\\ \today
\anotfoot{ *}{On leave from Grup de F\'{\i}sica Te\`orica and IFAE,
Universitat Aut\`onoma de Barcelona, Barcelona, Spain.
peris@surya11.cern.ch}
\anotfootnb{**}{Also at Institut de F\'{\i}sica Corpuscular (IFIC),
Val\`encia, Spain. santamar@goya.ific.uv.es}

\end{titlepage}
\setcounter{page}{1}

\mysection{Introduction}

Electroweak (EW)
radiative corrections are presently achieving an
extremely high degree of
sophistication and complexity. Indeed, after the pioneering one-loop
calculations of Marciano and Sirlin \cite{marciano-sirlin}
and the two-loop ones of
van der Bij and Veltman \cite{vanderBij-Veltman}, and because of the
high-precision experiments recently performed
at LEP and the SLC \cite{Miquel}, there is a clear need for increasingly
higher-order calculations;
even if only
for assessing the size of the theoretical error when comparing to the
experiment. Currently two-loop EW corrections (pure or mixed with
QCD) are being analyzed rather systematically \cite{Kniehl-report} and
, sometimes, even up to three loops are being accomplished \cite{Tarasov-rho}.
Needless to say these calculations are
extremely complicated
and usually heavily rely on the use of the computer.
In this paper we would like to point out that in some
situations thinking in terms of
effective field theories (EFTs) \cite{efts,efts2} can help much in this
development.

Built as a systematic approximation scheme for problems with widely
separated scales \cite{Hall}, EFTs organize the calculation
in a transparent
way dealing with
one scale at a time and clearly separating the physics of the
ultraviolet from the physics of the infrared. They are based on the
observation that, instead of obtaining the full answer and then take
the appropriate interesting limits, a more efficient strategy  consists
in taking the limit first, whereby considerably reducing the amount of
complexity one has to deal with, right from the start.
For this kind of problems EFTs are never more
complicated than the actual loopwise perturbative
calculation and in some
specific cases they may even be more advantageous, even
able to render an extremely
complicated calculation something very simple.

By EFT we specifically mean the systematic construction of the effective
Lagrangian that results when a heavy particle is integrated out. The
procedure goes as follows \cite{efts,efts2}. Let us
imagine we are interested in studying the
physics at an energy scale $E_0$. Starting at a scale $\mu>>E_0$ one uses
the powerful machinery of the renormalization group equations (RGE's) to
scale the initial Lagrangian from the scale $\mu$ down to the energy $E_0$
one is interested in. If in doing so one encounters a certain particle with
mass $m$, one must integrate this particle out and find the corresponding
matching conditions so that the physics below and above the scale $\mu=m$
(that is to say the physics described by the Lagrangian with and without the
heavy particle in question) is the same. This is technically achieved by
equating
the one-particle irreducible Green functions (with respect to the other
light fields) in both theories to a certain order in inverse powers of the
heavy mass $m$\footnote{One could also match S-matrix elements.}. This
usually requires the introduction of local counterterms \cite{Witten} in the
effective Lagrangian for $\mu<m$. Once this is done, one keeps using the
RGE's
 until the energy $E_0$ is reached. If another particle's threshold is
crossed, the above matching has to be performed again. All this procedure is
most efficiently carried out by using the \msb renormalization scheme
where the RGE's are mass independent and can be gotten directly from
the $1/\epsilon$ poles of dimensional regularization.

In this work we would like to apply this technique by concentrating on
the QCD corrections to the
large-$m_t$ EW contributions {\it specific} to the $Zb\bar b$ vertex.
Other corrections common to the other fermions originate from vacuum
polarization and have been already studied in \cite{Georgi-Cohen-Grinstein,
Grinstein-Wang, Peris};
therefore we shall not consider them. The decay width $\zbb$ can be
written as \cite{zbb-general,RQCD}

\begin{equation}
\label{one}
\Gamma(Z\to b \bar{b})= \N \ \G \ \ \rho \ \ R_{QCD}\  R_{QED} \
\ [A^2 + V^2]\ \ ,
\end{equation}
with
\begin{equation}
\label{two}
A=1 + {1\over 2} \Delta \rho^{vertex}\qquad ; \qquad
V=1 + {1\over 2} \Delta \rho^{vertex}- {4\over 3} \kappa s_0^2\ \ ,
\end{equation}
\begin{equation}
\label{three}
\kappa \approx  1 - {c^2\over c^2-s^2}\  \Delta\rho \ + {g^2\over \4pi^2}\
\ {1\over 6\ (c^2-s^2)}\log{M_W^2\over \mt^2}\
\left(1+{\a(\m)\over \pi}\right)\ \ ,
\end{equation}
\begin{equation}
\label{four}
\rho=1 + \Delta \rho \qquad ; \qquad
\Delta \rho\approx {3\over \4pi^2} \mt^2(m_t) \gf \sqrt 2
\left(1- {\a(\mu)\over \pi}\ \ {2\over 9}\ (\pi^2-9)\right) \ \ ,
\end{equation}
and
\begin{equation}
\label{five}
s_0^2={1\over 2}\left(1- \sqrt{1- {4\pi \alpha(M_Z)\over \sqrt 2 \gf
M_Z^2}}\right)\ \ ,
\end{equation}
where
\begin{equation}
\label{six}
R_{QCD}\approx 1 + {\a(\m)\over \pi}\quad ; \quad R_{QED}\approx 1 +
{\alpha(\m)\over 12 \pi} \ \ ,
\end{equation}
$$
\Delta \rho^{vertex}\approx  -{4 \mt^2(\mt) \gf\sqrt 2\over
\4pi^2}\left(1 -  {\a(\m)\over \pi} \ {\pi^2 - 8\over 3}\right)+
\qquad \qquad
$$
\begin{equation}
\label{seven}
\qquad \qquad +{g^2\over \4pi^2}\ \ \log
\left({M^2_W\over \mt^2}\right)\  \left({8\over 3}+
{1\over 6 c^2}\right)\left(
1+ \ {\cal C} \ {\a(\mu)\over \pi}\right) \ \ .
\end{equation}
It is more natural, when using the EFT language, to employ the running \msb
$\mt(\mu=\mt)$ rather than $\mt(pole)$ and this is what we have done in the
previous expressions.

What is the natural scale for $\m$ in these contributions?
As explained in refs.~\cite{Sirlin-Kniehl,Grinstein-Wang,Peris} the $\a(\m)$
appearing in $\Delta \rho$ of eq. (\ref{four}) is to be interpreted as
$\a(\m \approx m_t)$ because it originates at
the matching between the full theory with
top and the effective field theory
without top at the scale $\m=m_t$. As discussed in ref. ~\cite{Sirlin2}, it
turns out this
even encompasses most of the ${\cal O}(\a^2)$ contributions. On the
contrary, as explained in ref.~\cite{Peris}, the logarithmic term of
eq.~(\ref{three}) comes from the running of
the effective Lagrangian from the
scale $\m=m_t$ down to
the scale $\m=M_W\simeq M_Z$ and consequently does
not probe a single $\m$ scale but
rather integrates over the whole range. The
result of this integration leads to the substitution
\footnote{For small values of $\a$ of course the next two
expressions coincide.}

\begin{equation}
\label{eight}
\log{M_W^2\over m_t^2} \ \left(1 + {\a(\m)\over \pi}\right)\rightarrow
\log{M_W^2\over m_t^2}\ +\ \log\left({\a(M_W)\over
\a(m_t)}\right)^{-4/\beta_0}
\end{equation}
in eq. ~(\ref{three}); here $\beta_0\equiv 11- 2 n_f/3=23/3$ is
the $\beta$ function of the QCD
coupling constant $\a(\m)$ for $n_f=5$ flavors:

\begin{equation}
\label{nine}
{d\a\over dt}= - \ {\beta_0\over \4pi}\ \a^2(t)\quad ; \quad t\equiv \log
\mu^2\ \ .
\end{equation}
Equation (\ref{eight}) actually resums all the QCD leading logarithms,
i.e. all terms of the form $\a^n \log^n$.

In this paper we shall describe an effective field theory calculation of the
physical process $Z\rightarrow b \bar b$. As a result we shall obtain the
value of the coefficient ${\cal C}$ in eq. (\ref{seven}). This coefficient
has also been recently obtained in ref.~\cite{Kwiatkowski-Steinhauser}
and our result agrees with theirs. Moreover, our construction of the EFT
will also yield the value for the natural scale $\mu$ that appears in the
different terms of eqs. (\ref{six}),(\ref{seven}).

We think that our discussion in terms of
effective Lagrangians is a good guide
for
dealing with questions of this sort and also for computing
things like ``the QCD
corrections to the $\log m_t$ term'' for large $m_t$,
i.e. the coefficient ${\cal C}$. As we will see,
one could even resum, if necessary, the leading
logarithms. However one
should keep in mind that in the real
world the top mass is not that large
with respect to the $Z$ and, therefore,
one should expect sizeable corrections
to the simple case $m_t\to \infty$.

\mysection{Case without QCD corrections}

Integrating the top quark out affects the coupling
to the $W$ and $Z$ gauge bosons of every lighter fermion
through vacuum polarization. Moreover it also
affects specifically the coupling of the
bottom quark to the $Z$ boson; an
effect that is not felt by any
other fermion. To set the stage for the
QCD corrections of the next section we shall now review, in an effective
field theory language, how this comes about. The standard strategy is
the following:
\begin{enumerate}
\item Matching the effective theory to the full theory at $m_t$.
\item Running the effective Lagrangian from $m_t$ down to $M_Z$.
\item Calculating matrix elements with the effective Lagrangian at the
scale $M_Z$.
\end{enumerate}

The integration of the top quark is done in
several steps. Firstly, at tree
level, there is the contribution given by the diagram of fig.~\ref{zbb1}.
This contribution gives rise to an effective operator
that is suppressed by two
inverse powers of the top mass. We shall consistently
neglect this type of
contributions since they can never give rise to the terms we are
interested
in, i.e. eq. (\ref{seven}). This is the only contribution in the
unitary gauge, which is the one we shall
employ in this work. In any other
gauge other effective operators arise
because the would-be Nambu--Goldstone
bosons couple proportionally to the top
mass and may compensate the $m_t^2$
factor in the denominator (see for example
the diagram of fig.~\ref{zbb2}). This is a
welcome simplification, most notably when
QCD corrections will be considered
in the next section.

Since at this order the matching condition turns out to be trivial,
the effective Lagrangian below the top quark mass
looks exactly the same as
the full standard model Lagrangian except that the
top is absent; e.g. there
is no $tbW$ vertex in this Lagrangian
:

$$
{\cal L}=\bar b\ i\Dslash\ b - \frac{1}{2}\gt^b(\m)\
\bar b\ \Zslash\ P_L\ b +
{1\over 3} c_V^b(\m)\ \bar b\ \Zslash\ b +
$$
\beq
\label{one2}
+\bar e\ i \Dslash\ e -\frac{1}{2} \gt(\m)\ \bar e\ \Zslash\ P_L\ e +
c_V(\m) \bar e\ \Zslash\ e + \frac{c_+(\m)}{\sqrt 2}
\left(\bar e\ \Wslash\
P_L\ \nu + h.c.\right)
\eeq
where $P_L$ is the lefthanded projector and
$\Dslash$ stands only for the QED covariant derivative since $\a=0$ in
this section. The
$c(\m)$'s of the electron are actually
common to all the fermions but the
bottom quark. For instance, the $Z\nu \bar \nu$ would be $+c_L(\m)/2$
since the neutrino has no vector coupling $c_V(\m)$.
We also take
all the fermions but the top as massless and assume that $M_H\simeq
M_{Z,W}$ to avoid unnecessary complications in the form of terms
$\log (M_H/M_Z)$.
Notice
that we have decomposed the $Zf\bar f$ vertex in terms of a lefthanded and
vector couplings instead of the more conventional left and righthanded, or
vector and axial counterparts. The matching conditions at this order yield
Standard Model tree-level values for the effective couplings:
\beq
\label{two1}
\gt^b(m_t)=\frac{g}{c}\quad ;\quad c_V^b(m_t)=\frac{g}{c}\ s^2\ ;
\eeq
\beq
\label{two2}
c_L(m_t)=\frac{g}{c}\quad ;\quad  c_V(m_t)=\frac{g}{c}\ s^2\quad ;
\quad c_+(m_t)=g\ \ .
\eeq

We can select  the non-universal part of the $Zb\bar b$ vertex by comparing
the $\gt^b(\m)$ coupling on shell with the analogous coupling for the
electron $\gt(\m)$ at the scale $\m\sim M_Z\sim M_W\equiv M$. One
defines\footnote{This ratio is called $1+\epsilon_b$ in
ref.~\cite{Altarelli-Barbieri-Caravaglios-NPB405}}
\beq
\label{alta}
1+\frac{1}{2} \Delta \rho^{vertex} = \frac{\gt^b(M)}{\gt(M)}\ \ .
\eeq

In order to make contact with the physics at the scale $\m=M$, one has to
scale the Lagrangian (\ref{one2}) down to this particular $\m$.  Since we
are only interested in one-loop electroweak contributions, the RGEs that
govern this scaling will be computed using the lowest order in the
electroweak couplings, i.e. the tree level. In this process of scaling,
$\gt^b(\m)$ and $\gt(\m)$ run differently. The calculation can be
done by setting the external particles on shell. This is actually a subtle
point. If one were to compute off-shell Green functions one would discover
that extra structures other than those appearing in the operators of eq.
(\ref{one2}) are generated in the running. This has nothing to do with the
effective field theory construction but is a reflection
of the known fact that the Standard Model is not manifestly renormalizable in
the unitary gauge at the level of off-shell Green functions. Only when
S-matrix elements are taken can one renormalize it
\cite{Fujikawa-Lee-Sanda-Pokorski-book11}.
Of course this is all we need since in the
physical process $Z\rightarrow b \bar b$ all particles are on shell,
 so that
the physical amplitude is governed by the coefficient $c^b_L(M)$.

For $\gt^b(\m)$ the running is
given by the $1/\epsilon$ poles of the
diagrams of fig.~\ref{zbb3}. One can see, however, that the diagrams of
figs.~\ref{zbb3}a,b do not contribute to the
running of $\gt^b(\m)$ since they are
actually finite (this is akin to the Ward--Takahashi
identity relating the
vertex to the wave function renormalization in QED) and the only
contribution comes from fig.~\ref{zbb3}c. Therefore
\beq
\label{three1}
\frac{d\gt^b(t)}{dt}= \mathrm{fig.~\ref{zbb3}c}\ \ .
\eeq
However things look differently for
$\gt(\m)$ since {\it there is} an $e\nu
W$ vertex. The running of $\gt(\m)$ is
given by the diagrams of fig.~\ref{zbb4}. As
before, the diagrams \ref{zbb4}d,e do not contribute,
and an explicit calculation in the unitary gauge  for on-mass-shell matrix
elements shows that
the diagrams of fig.~\ref{zbb4}b,c are each finite.
Therefore the only
contribution is that of figs.~\ref{zbb4}a and \ref{zbb4}f, and one finds
\beq
\label{four1}
\frac{d\gt(t)}{dt}=-\ \frac{g}{c}\
\frac{g^2}{\4pi^2}\ \left(\frac{4}{3} + \frac{1}{12
c^2}\right)\ + \mathrm{fig.~\ref{zbb4}f}\ \ ,
\eeq
where the relation $M_Z^2 c^2=M_W^2$ has been used.
One can now trivially integrate eqs. (\ref{three1})-(\ref{four1})
with respect
to t between $\log m_t^2$ and $\log M^2$ to find

\beq
\label{five1}
\gt^b(M) \simeq \gt^b(m_t) + (\mathrm{fig.~\ref{zbb3}c}) \log \frac{M^2}{m_t^2}
\eeq
\beq
\label{five2}
\gt(M) \simeq \gt(m_t) - \frac{g}{c}\
\frac{g^2}{\4pi^2}\ \left(\frac{4}{3} + \frac{1}{12
c^2}\right) \log\frac{M^2}{m_t^2} + (\mathrm{fig.~\ref{zbb4}f})
\log\frac{M^2}{m_t^2}\ \ .
\eeq
This yields
\beq
\label{six1}
\frac{\gt^b(M)}{\gt(M)} \simeq  \frac{\gt^b(m_t)}{\gt(m_t)}\ \left(1 +
\frac{g^2}{\4pi^2}\left(\frac{4}{3} + \frac{1}{12 c^2}\right) \log
\frac{M^2}{m_t^2}\right)\ \ ,
\eeq
since the contribution of figs.~\ref{zbb3}c and \ref{zbb4}f
cancel each other out in this
ratio. The tree-level matching conditions of eqs. (\ref{two1}),
(\ref{two2}) lead to
$\gt^b(\mt)/\gt(m_t)=1$ and, therefore,
in this approximation, one finds the
right logarithmic piece of eq. (\ref{seven}) but the power-like one that
goes like $m_t^2$ is still missing.
To obtain it one has to go to the next
order, i.e. do a one-loop matching. In order to do this one has to
consider the integration of the
top quark to one loop. This not only modifies the boundary
conditions of eq. (\ref{two1}),(\ref{two2}) but
also produces new nontrivial
operators besides those of eq. (\ref{one2}). The relevant diagrams are
depicted in fig.~\ref{zbb5}. Then the part
of the Lagrangian involving the bottom quark interactions
reads\footnote{Because of the diagram of fig.~\ref{zbb5}a
there is
a wave function renormalization
factor. ${\cal L}_4$ only follows after making
the field redefinition that renders the kinetic term in a standard form.}
\beq
\label{seven1}
{\cal L}={\cal L}_4 + {\cal L}_6 \ \ ,
\eeq
\beq
\label{seven2}
{\cal L}_4= \bar b\ i\Dslash\ b - \frac{1}{2}\gt^b(\m)\
\bar b\ \Zslash\ P_L\ b +
 \frac{1}{3}\ c_V^b(\m)\ \bar b\ \Zslash\ b
\eeq
and
\beq
\label{seven3}
{\cal L}_6=\frac{1}{\Lambda_F^2}\ \sum_i\ c_i(\m) {\cal O}_i \ \ ,
\eeq
where $\Lambda_F=4 \pi v$,
$v=(\sqrt 2 G_F)^{-1/2}= 246$~GeV and  the ${\cal O}_i$'s are a set of
dimension six operators involving the (lefthanded)
bottom quark and three derivatives; or
the (lefthanded) bottom quark, the $Z$ and two
derivatives\footnote{See next section for
more discussion.}. They arise from the longitudinal part of the $W$
propagators. This is why the scale $\Lambda_F$ appears: it is the
combination of the ordinary $1/m_t^2$ suppression of any six dimensional
operator in an effective field theory and the fact that the would-be
Nambu--Goldstone bosons couple proportionally to the top mass. There are
also other operators generated at this stage, like for instance a
Wess-Zumino term (which ensures in the theory without top the
cancellation of anomalies that occurs in the theory with top) or a
four bottom-quark operator (which comes from box diagrams with the top
flowing in the loop). However they can only affect the $Zb\bar b$
vertex with contributions that are two-loop electroweak, i.e.
subleading with respect to eq. (\ref{seven}).
Still higher
dimensional operators may exist but they
are truly suppressed by inverse powers of the top
mass.

It is easy to convince oneself that, for $\a=0$, one can forget
about ${\cal
L}_6$. Firstly, we are only interested in considering matrix elements
of ${\cal L}_6$ at tree level since ${\cal L}_6$ itself has been
generated at one electroweak loop.
With a massless bottom,
dimensional analysis leaves $M_Z$ as the only scale to compensate the scale
$\Lambda_F$ in eq. (\ref{seven3}). This is an effect of ${\cal O}(g^2)$ in
the $Zb \bar b$ vertex, i.e. without the $\log m_t$ enhancement and
therefore is subleading with respect to those of eq. (\ref{seven}).

However, this is not the whole story. The longitudinal part
of the $W$ boson
propagator also produces a contribution to
the matching condition of the
$\gt^b(\m)$ coupling of eq.
(\ref{seven2}) and modifies the first of the boundary conditions in eq.
(\ref{two1}). An explicit calculation yields the following result
\beq
\label{ten2}
\frac{\gt^b(m_t)}{\gt(m_t)}= 1 -2 \ \frac{m_t^2}{(4\pi v)^2}\ \ ,
\eeq
since $c_L(m_t)=g/c$ remains unchanged. Inserting this into
\eq{six1} one obtains the desired final result:
eq. (\ref{six1}) one
\beq
\label{ten1}
\frac{\gt^b(M)}{\gt(M)} \simeq 1 -2 \ \frac{m_t^2}{(4\pi v)^2}+
\frac{g^2}{\4pi^2}\left(\frac{4}{3} + \frac{1}{12
c^2}\right)\log\frac{M^2}{m_t^2}\ \ ,
\eeq
i.e. eq. (\ref{seven}) with $\a=0$. For the effective
field theory aficionado eq. (\ref{ten1}) is somewhat
unconventional in that it mixes matching (i.e. the $m_t^2$ term)
with running (i.e. the logarithm) both at one loop. The reason is of course
the non-decoupling of the top quark which enhances the one-loop matching
contribution with the sizeable $m_t^2$ term. The rest of the one-loop
matching {\it is} of ${\cal O}(\alpha)$ and therefore subleading.

Clearly the effect of integrating the top quark out affects only the
lefthanded projection of the bottom-quark field, $c_L^b(\m)$, but
leaves untouched the coefficient $c_V^b(\m)$. This coefficient not only
equals $c_V(\m)$ at $\m=m_t$ (eqs. (\ref{two1}-\ref{two2}))
but also runs with
$\m$ as $c_V(\m)$ does. In other words, $c_V^b(\m)=c_V(\m)$. Of course
this fact will be unaffected by QCD corrections.
\mysection{QCD corrections}

As discussed in the previous section, the unitary gauge has the
advantage that, at the tree
level,  the matching corrections that appear when one integrates out
the top quark
are suppressed by two inverse powers of the top quark mass (fig.
{}~\ref{zbb1}). Since ultimately
this fact is due to dimensional analysis, it cannot change once QCD is
switched on and one-loop $\a$ corrections to the diagram of fig.~\ref{zbb1}
are also considered in the matching conditions. Therefore, with our choice
of the unitary gauge, dimensional analysis still dictates that these
corrections are also suppressed by two inverse powers of the top mass when
$\a \not= 0$.
Since in the matching conditions one has to calculate not only the
divergent parts but also the finite pieces, the fact that the unitary
gauge does away with these particular matching conditions altogether is
a major simplification and justifies our choice of this gauge.

Not all the matching corrections disappear, however. The diagrams of
fig.~\ref{zbb6} do give rise to new dimension
six operators. Since no gluon loop appears in these diagrams, these
operators are explicitly QCD gauge invariant and, in fact, together
with the diagrams of fig.~\ref{zbb5} they are
none other than the Lagrangian ${\cal L}$ of eqs.
(\ref{seven1})-(\ref{seven3})
with the only change that, now, every
derivative has been appropriately promoted to a QCD covariant
derivative. It is important to notice that both sets of diagrams
(fig.~\ref{zbb5} and fig.~\ref{zbb6}) are necessary in order to resolve the
ambiguities one encounters when trying to implement this promotion to
covariant derivatives.

In principle one should now calculate how all these operators mix back
into the $Zb\bar b$ operator of eq. (\ref{seven2}) and make the
coefficients $c^b_{L,V}(\m)$ evolve with $\m$ as one runs from $m_t$
down to $M_Z$. However, one notices that since we are only interested
in on-shell $Z$ gauge bosons we can use the free-field equations of
motion for the $Z$ to trade derivatives by $Z$ masses. Moreover, although the
$b$-quark field is not on-shell and will be closed in loops, one is also
free to use the {\it full} field equations of motion for the $b$ quark as
a means of defining and simplifying the operator basis \cite{Arzt}.

In this way, a clever
use of the equations of motion helps us get rid of most of the
operator structures that are generated in the matching and leaves us
with only three (in principle) relevant operators. These
are\footnote{One could still use the equations of motion for the gluon
field but we found more convenient not to do so.}{}$^,$\footnote{There are two
Hermitian linear combinations of $\op_2$ that must be considered.}
\begin{eqnarray}
\label{arca1}
\op_1 &=& \bbl \gamma^\nu \frac{\lambda^A}{2}
\bl \ g_s\ D^\mu G^A_{\mu \nu}\nonumber\\
\op_2 &=& \frac{g}{c}\bbl \sigma^{\mu\nu} \sla{Z} \frac{\lambda^A}{2}\bl
g_s G^A_{\mu\nu}\\
\op_3 &=& \ \frac{g}{c}\bbl \gamma^\mu \frac{\lambda^A}{2}  \bl Z^\nu
g_s G^A_{\mu\nu}\ \ .
\end{eqnarray}
Of course these operators only affect the running of $c^b_L(\m)$ and not
of $c_V^b(\m)$.

For instance an operator like  $\op_4 = i\bbl \db^3 \bl $  will give
rise, upon integration by parts and
use of the equations of motion for both $b$
fields, to another
operator with two $Z$'s which can only mix back into the $Zb \bar b$
operators of eq. (\ref{seven2}) by closing one $Z$ in a loop, i.e. at the
two-electroweak-loop level since $\op_4$ originated already at one
electroweak loop, and therefore we can neglect it. One can check the fact
that the $\op_4$ can be neglected  by computing its contribution to the
diagrams of figs.~\ref{zbb7}a and ~\ref{zbb7}b and seeing that they cancel
each other. Throughout this work we shall
always use the Feynman gauge propagator for the gluon.

An explicit straightforward evaluation of
the diagrams of fig.~\ref{zbb6}a yields
for the coefficient $c_1(\m)$ accompanying the operator $\op_1$ the
value
\beq
\label{arca2}
c_1(m_t)=-\frac{7}{18} \ \ .
\eeq

In principle we should also compute $c_{2,3}(m_t)$ for the
corresponding operators $\op_{2,3}$. However, for our purposes this is
totally unnecessary. A moment's thought reveals that $\op_{2,3}$ can
never mix back into the $Zb \bar b$ operators of eq. (\ref{seven2}),
where the three particles are on shell, because closing the gluon line
in a loop will produce --through a $\pslash$ or a $p^2$--
a bottom mass which we have taken to be zero.

In order to make contact with the physics at the scale $\m \simeq M$
one has to find how $c_L^b(\m)$ scales with $\m$. What is the RGE
governing the running of the coefficient $c_L^b(\m)$ now that $\a$
corrections are included? For $\a=0$ we know that the answer is given
by eq. (\ref{three1}). Not much is changed when $\a\not= 0$. In particular,
the cancellation of the diagrams of figs.~\ref{zbb3}a and ~\ref{zbb3}b
still takes place, even with QCD corrections. First of
all there is the diagram of fig.~\ref{zbb3}c (now of course including
$\a$corrections in the $Z$ vacuum polarization). Notice that there can be
no gluon attaching the vacuum polarization to the external bottom-quark
line to order $\a$. Secondly there is the contribution of the
coefficients $c_i(\m)$ of ${\cal L}_6$ to the RGE for $c_L^b(\m)$.
However, as we discussed above, only $c_1(\m)$ needs to be considered;
the relevant diagrams are those of fig.~\ref{zbb7}b. Notice also in this
regard that there
is no wave function renormalization due to $\op_1$ because of the
masslessness of the bottom quark.
Finally notice
that $c^b_L(\m)$ does not renormalize itself through $\a$ corrections
because the operator $\bbl \Zslash \bl $ behaves like a conserved current
under QCD.

Consequently, gathering all the pieces, one obtains that
\beq
\label{arca3}
\frac{dc_L^b(t)}{dt}=\mathrm{ fig.~\ref{zbb3}c}\
+\ \frac{g}{c}\ \frac{g^2}{\4pi^2}\ \gamma_1\ c_1(t) \frac{\a(t)}{\pi}
   \ \ \ .
\eeq

We have explicitly checked that, as we argued before,
the contributions of diagrams
of figs.~\ref{zbb8}a and \ref{zbb8}b cancel out when the vertices
obtained from figs.~\ref{zbb5}b and \ref{zbb6}b are inserted
thus confirming that the only running, besides
the vacuum polarization diagram of fig.~\ref{zbb3}c, comes from the penguin
operator $\op_1$, so that eq. (\ref{arca3}) indeed follows. We obtain
the following value for the coefficient $\gamma_1$:

\beq
\label{gamma1}
\gamma_1=-\frac{1}{9 c^2}\left(1-\frac{2}{3} s^2\right)\ \ .
\eeq
Since $\op_1$ only involves the lefthanded bottom quark it is clear why the
coefficient $\gamma_1$ turns out to be proportional to the lefthanded bottom
coupling to the $Z$, i.e. the combination $1-\frac{2}{3}s^2$.

Now we would like to integrate eq. (\ref{arca3}). In first
approximation, one may take $\a(t)$ and $c_1(t)$ as constants
independent of $t$, i.e. $\a(\m)\simeq \a(m_t)\simeq \a(M)\equiv \a$
and $c_1(\m)\simeq c_1(m_t)\simeq c_1(M)\equiv c_1 = -7/18$. The
integration over $t$ between $\log m_t^2$ and $\log M^2$ gives

\beq
\label{arca4}
c_L^b(M)\simeq c_L^b(m_t) + \mathrm{fig.~\ref{zbb3}c}
\ \log \frac{M^2}{m_t^2} +
\frac{g}{c} \frac{g^2}{\4pi^2}\ \gamma_1 \ c_1\ \frac{\a}{\pi} \log
\frac{M^2}{m_t^2} \ \ .
\eeq
It is in principle possible to improve on this approximation by
considering the $\m$-dependence of $\a(\m)$ and $c_1(\m)$ in eq.
(\ref{arca3}). The
$\m$-dependence of $\a(\m)$ is given by the usual one-loop $\beta$
function. However the $\m$-dependence of $c_1(\m)$ is more complicated
to obtain because it requires performing a complete operator mixing
analysis of the penguin operator along the lines of, for instance,
the work carried
out in the studies of $b\rightarrow s\gamma$ or $K^0$-$\bar K^0$ mixing
\cite{Martinelli-altres} from where most of the results
could be taken over to our case. However, the fact that
$\gamma_1 c_1(m_t)=\frac{7}{162 c^2}\ (1-\frac{2}{3}\ s^2)\approx 0.05$
turns out to be so small renders this improvement moot and we
shall content ourselves with eq. (\ref{arca4}) as it is. As we shall see
later on, there are other sources of QCD corrections that are numerically
more important.

Of course leptons do not have strong interactions and for them the
answer for the running of $c_L(\m)$ is still eq. (\ref{five2}) (notice
however that fig.~\ref{zbb4}f will now have some $\a$-dependence inside the
vacuum polarization, just as fig.~\ref{zbb3}c does). Consequently one obtains
from eqs. (\ref{arca4}),(\ref{five2}):
\beq
\label{arca5}
\frac{c_L^b(M)}{c_L(M)} \approx \frac{c_L^b(m_t)}{c_L(\mt)} \left[1 +
\frac{g^2}{\4pi^2}
\left(\frac{4}{3}+\frac{1}{12c^2}\right)\log\frac{M^2}{m_t^2}+
\frac{g^2}{\4pi^2} \ \gamma_1 \ c_1 \frac{\a}{\pi}\log\frac{M^2}{m_t^2}
\right] \ \ ,
\eeq
since, also in this case, the contributions from figs.~\ref{zbb3}c and
\ref{zbb4}f cancel
each other out in this ratio. This fixes the coefficient ${\cal C}$ in
eq. (\ref{seven}) to be

\beq
\label{arcaarca}
{\cal C}= 2\ \gamma_1 \ c_1\ \left(\frac{8}{3}+\frac{1}{6 c^2}\right)^{-1}
\approx 0.03\ \ .
\eeq

We would like to mention at this point that all of the above calculations
(included those in the previous section without QCD) have been repeated
using the Landau gauge for the $W$ propagator with exactly the same
results. In particular we would like to point out that in this gauge
the coefficient $c_1(\mt)$ comes entirely from fig.~\ref{zbb6}a with the
$W$ replaced by a Nambu--Goldstone boson.
This agreement has as a consequence the vanishing of the
contribution to $c^b_L(M)$ of the QCD corrections to the matching
conditions of operators with external Nambu--Goldstone bosons like for
instance the one in fig.~\ref{zbb2}, which are present
in this gauge. Of course in the unitary gauge they trivially disappear.

The ratio
$c_L^b(m_t)/c_L(m_t)$ is still given by eq. (\ref{ten2}) in this
approximation. One could now consider $\a$ corrections to eq.
(\ref{ten2}). This is a hard two-loop calculation of the matching
conditions in the presence of QCD when the top is integrated out. Here
the effective field theory language does not help much and the
calculation has to be done. Fortunately the result is already available
in the literature \cite{match-mt-2loops}. Translated into our
context it amounts to
\beq
\label{arca6}
\frac{c_L^b(m_t)}{c_L(m_t)} = 1 - 2\ \frac{m_t^2(\mt)}{(4\pi v)^2}\left[1 -
\frac{\a(\mt)}{\pi}\ \left( \frac{\pi^2-8}{3}\right)\right]\ \ .
\eeq

Again, what the EFT does tell us is that the $\m$ scale of $\a(\m)$
in this equation  has to be $m_t$ since it originates at the matching
condition when the top is integrated out.\footnote{Another advantage is
that matching conditions are free from infrared divergences\cite{efts},
which is a nice simplification. For
some more discussion on infrared divergences, see below.} Therefore we
get to eq. (\ref{seven}) with $\a(\m=m_t)$ in the $m_t^2$-dependent
term.

However this is not yet all. Up to now all the physics has been
described with RGEs (i.e. running) and their initial
conditions (i.e. matching)
which is only ultraviolet physics, and
no reference to infrared physics has
been made. For instance, where are the infrared divergences that appear
when a gluon is radiated off a bottom-quark leg? As we shall now see,
this physics is in
the matrix element for $Z\rightarrow b\bar b$. After
all, we have only obtained the effective Lagrangian
(\ref{seven1}),(\ref{seven3}) at the scale $\mu=M$; we still have to
compute the physical matrix element with it, and here is where all the
infrared physics takes place.

Indeed, when computing the matrix element for $Z\rightarrow b \bar b$
with the effective Lagrangian (\ref{seven1}),(\ref{seven3}) expressed in
terms of $c_{L,V}^b(\m)$ at $\m=M$, one has the contribution of the
diagrams of figs.~\ref{zbb9}a, \ref{zbb9}b,
where the $\otimes$ stands for the
effective vertices proportional to $c_{L,V}^b(M)$. These diagrams give
rise to infrared divergences. These divergences disappear in the
standard way once bremsstrahlung diagrams like those of fig.~\ref{zbb9}c
are (incoherently) added \cite{IR,bile}. Note that similar diagrams with
effective $g-b-b$ couplings are subleading and never give rise to
corrections of the form of eq. (\ref{seven}).

As is well known \cite{RQCD}, the net result of all this
(a similar calculation can be performed for the QED corrections)
is the appearance of the
factors $R_{QCD}$ and $R_{QED}$ of eqs. (\ref{one}),(\ref{six}),
where $b$-quark mass effects can also be included
\cite{djouadi-chetyrkin,bile} if needed.

The EFT
technology adds to this the choice of scale for $\m$, namely $\m=M$, in
these factors\footnote{This has been previously suggested by D. Bardin
(private communication).}:
\beq
\label{arca7}
R_{QCD}\simeq 1 + \frac{\a(M)}{\pi}\quad ; \quad R_{QED} \simeq 1 +
\frac{\alpha(M)}{12\pi}\ \ ,
\eeq
and naturally leads to the factorized expression (\ref{one})-(\ref{seven})
(see the Appendix),
with the value of ${\cal C}$ given by eq. (\ref{arcaarca}). As stated in the
introduction, our result agrees with that of
ref.~\cite{Kwiatkowski-Steinhauser}. Since the
``intrinsic'' $\a$ contribution of $\Delta \rho^{vertex}$ is, due to the
smallness of the coefficient ${\cal C}$, much less important than that
of $R_{QCD}$ one sees that the QCD corrections to the non-universal
$\log m_t$ piece of the $Zb \bar b$ vertex are, to a very good
approximation, of the form one-loop QCD ($m_t<<M_Z$) times one-loop
electroweak ($m_t>>M_Z$)\cite{Peris-unpublished}.

\section{Discussion and conclusions}

In this paper we have presented an effective field theory study
of the process $\zbb$ in the limit of large $\mt$  including
all the QCD corrections to the leading and the subleading contributions.
In particular we have explicitly shown that the QCD corrections to the
term $\log(\mt/M_Z)$ can be easily obtained in the EFT framework by
computing only a couple of one-loop diagrams in the limit of small
external momenta. We have also shown that the EFT framework answers
quite naturally the question of the renormalization
points to be used for the coupling constants
in the different terms.

In addition, it is important to remark that in the
EFT language all the physics
above $M$ is absorbed (in particular, all $\mt$ effects) in the
coefficients of the effective operators so that infrared physics is
relegated to the calculation of the physical process one is
interested in. With our effective Lagrangian
one could in principle compute any physical quantity, and not only
the $Z$ width, like for example jet production (i.e. where cuts are needed)
, forward--backward asymmetries, etc.
This is to be compared with more standard methods for computing radiative
corrections to the $Z$ width in which this width is extracted from the
imaginary part of the $Z$ self-energy to avoid problems
with infrared divergences. For instance, a more standard calculation
of the coefficient ${\cal C}$ \cite{Kwiatkowski-Steinhauser} requires
starting with the $Z$ self-energy at three loops (with all the complications
of the renormalization program at that order) to then compute its
imaginary part.
Extracting from this spatial asymmetries or jet rates, for instance, will be
much harder because it is not easy to limit to one's needs the entire
phase space.
The EFT calculation clearly separates ultraviolet from infrared physics and
as a consequence it is more flexible. And it is also simpler since, after
all, we never had to compute anything more complicated than a one-loop
diagram.

Of course, our results become more accurate as the top mass becomes larger.
In practice it is unlikely that the top quark be much heavier than, say,
$200$~GeV so due caution is recommended in the phenomenological use of eq.
(\ref{one}). In the lack of a (very hard !) full ${\cal
O}(g^2\a)$ calculation, this is the best one can offer. Furthermore, we
think it is interesting that at least there exists a limit (i.e. $m_t>>M_Z$)
where the various contributions are under full theoretical control.

\section*{Acknowledgements}
We would like to thank G. Martinelli for a helpful and enjoyable
discussion on the subject of this paper. S.P. would also like to thank
T. Mannel and M. Jamin for interesting discussions and K.G.
Chetyrkin, J.H. K\"uhn and M. Steinhauser for interesting conversations
during the Ringberg workshop "Perspectives for electroweak interactions in
$e^+e^-$ collisions", Munich, Feb. 1995, where part of this work was
presented. He would also like to thank B. Kniehl for the very good
organization of this workshop and for the kind invitation.

This work was  supported in part by
CICYT, Spain, under grants AEN93-0474 and AEN93-0234.

\vfill\eject

\appendix

\begin{center}{\Large\bf APPENDIX}\end{center}

\mysection{The width for $\zbb$}

In ref.~\cite{Peris} the effective couplings $g_3(\m), g_+(\m)$ and $g'(\m)$
were employed. The connection with $\cL(\m), c_V(\m)$ and $c_+(\m)$ is
$$
c_L(\m)=\frac{g_3(\m)}{c(\m)}\ ;
\ c_V(\m)=\frac{g_3(\m) s^2(\m)}{c(\m)}\
;\ c_+(\m)=g_+(\m)\ \ ,
$$
where $s^2(\m)=\sin^2\theta_W(\m)$ and $\tan\theta_W(\m)=
g'(\m)/g_3(\m)$.

According to our discussion, the width $Z\rightarrow b \bar b$ can be
computed from our Lagrangian (\ref{seven1}) as
$$
\Gamma \propto M_Z \ R_{QCD}\ R_{QED}\ \left[ V^2 + A^2\right]\ \ ,
$$
where $R_{QCD,QED}$ are given by eq (\ref{six}) with $\mu=M$, and
$$
V=\frac{1}{3}\ c_V^b(M) - \frac{1}{4}\ c_L^b(M)\ \ ,
$$
$$
A=\frac{1}{4}\ c_L^b(M)\ \ .
$$
Therefore
$$
V^2+A^2=\frac{1}{16}\left[\left(c_L^b(M)\right)^2 +
\left(c_L^b(M)-\frac{4}{3} c_V^b(M)\right)^2\right]\ .
$$

Since $c_V^b(\m)=c_V(\m)$ and $c_V(\m)=c_L(\m) s^2(\m)$ one obtains
that
$$
V^2+A^2=\frac{1}{16}\ \left(c_L(M)\right)^2\
\left[\left(\frac{c_L^b(M)}{c_L(M)}\right)^2 +
\left(\frac{c_L^b(M)}{c_L(M)}-\frac{4}{3} s^2(M)\right)^2\right]\ \ ,
$$
and, using (\ref{arca5})-(\ref{arca6}), one sees that
$$
\frac{c_L^b(M)}{c_L(M)}= 1 + \frac{\Delta\rho^{vertex}}{2}\ \ .
$$

 From ref.~\cite{Peris} we know that
$$
M_W^2 = \frac{c_+^2(M)}{4}\ v_+^2(M) = \frac{c_+^2(M)}{4}\ \left(\sqrt
2 G_F\right)^{-1}\ \ ,
$$
and
$$
M_Z^2 = \frac{c_L^2(M)}{4}\ \left(v_3^2(M) + 4 M_Z^2
\delta Z_{3Y} (M)\right)\ \ ,
$$
so that one can relate $c_L(M)$ to $M_Z$ and $G_F$. One finally obtains
eq. (\ref{one}) after noticing that
$$
s^2(M)= \kappa s_0^2\ \ ,
$$
with
$$
\kappa= 1-\frac{c^2}{c^2-s^2}\ \Delta\rho + \frac{g^2}{c^2-s^2} \ \delta
Z_{3Y}(M)\ \ ,
$$
and
$$
\delta Z_{3Y}(M)= \frac{1}{6 \4pi^2}\left[ \log \frac{M^2}{m_t^2} +
\log \left(\frac{\a(M)}{\a(m_t)}\right)^{-4/\beta_0} \right] \ \ .
$$
\vfil\eject

\vfil\eject
%
\def\mafigura#1#2#3#4{
  \begin{figure}[hbtp]
    \begin{center}
      \epsfxsize=#1 \leavevmode \epsffile{#2}
    \end{center}
    \caption{#3}
    \label{#4}
  \end{figure} }
\mafigura{6cm}{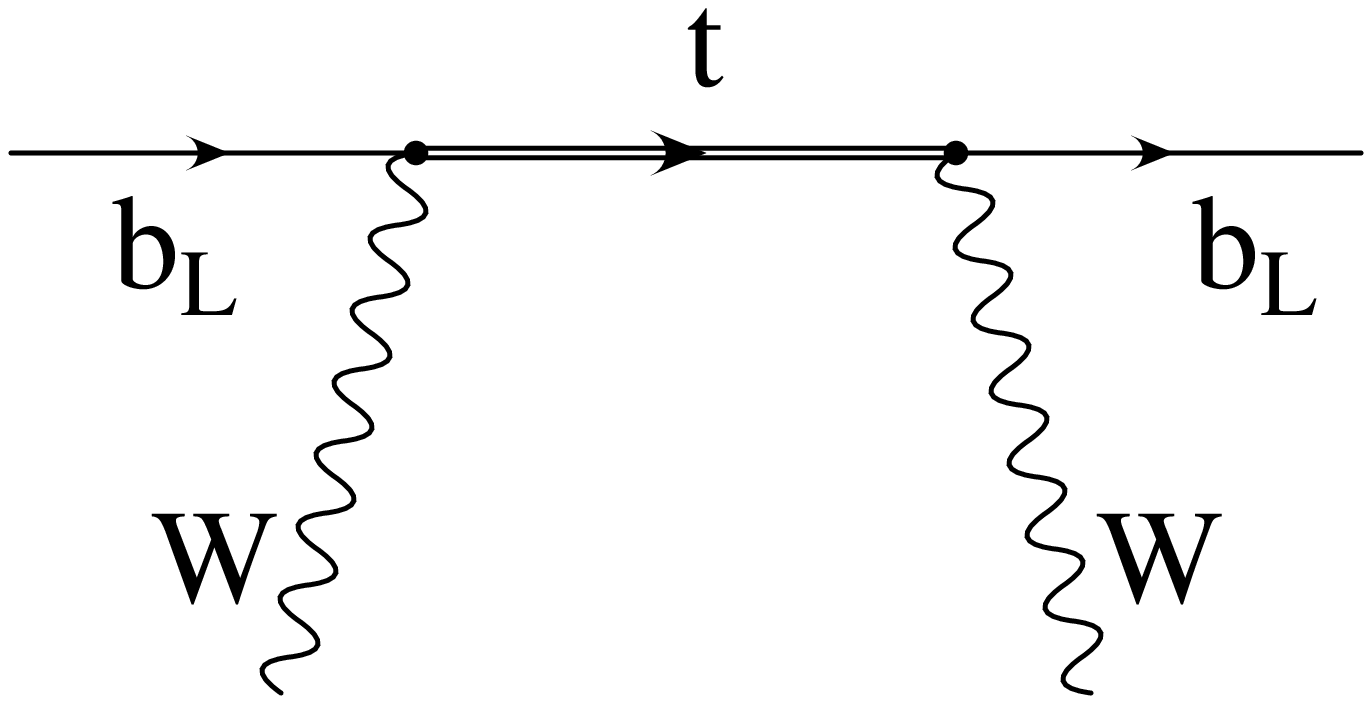}{Diagram contributing to the matching in the
unitary gauge. It is suppressed by $1/\mt^2$.}{zbb1}
\mafigura{13cm}{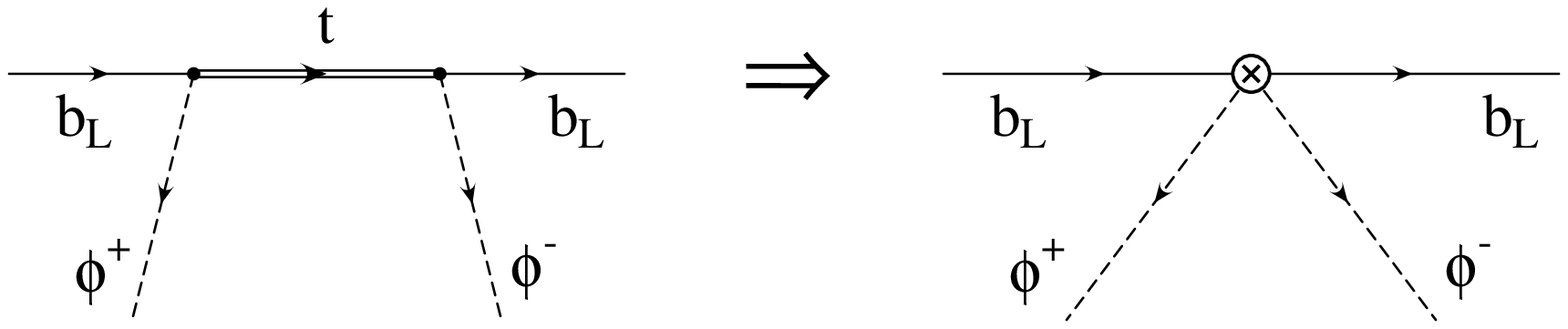}{Diagram contributing to the matching in
non-unitary gauges. It is not suppressed by $1/\mt^2$.}{zbb2}
\mafigura{16cm}{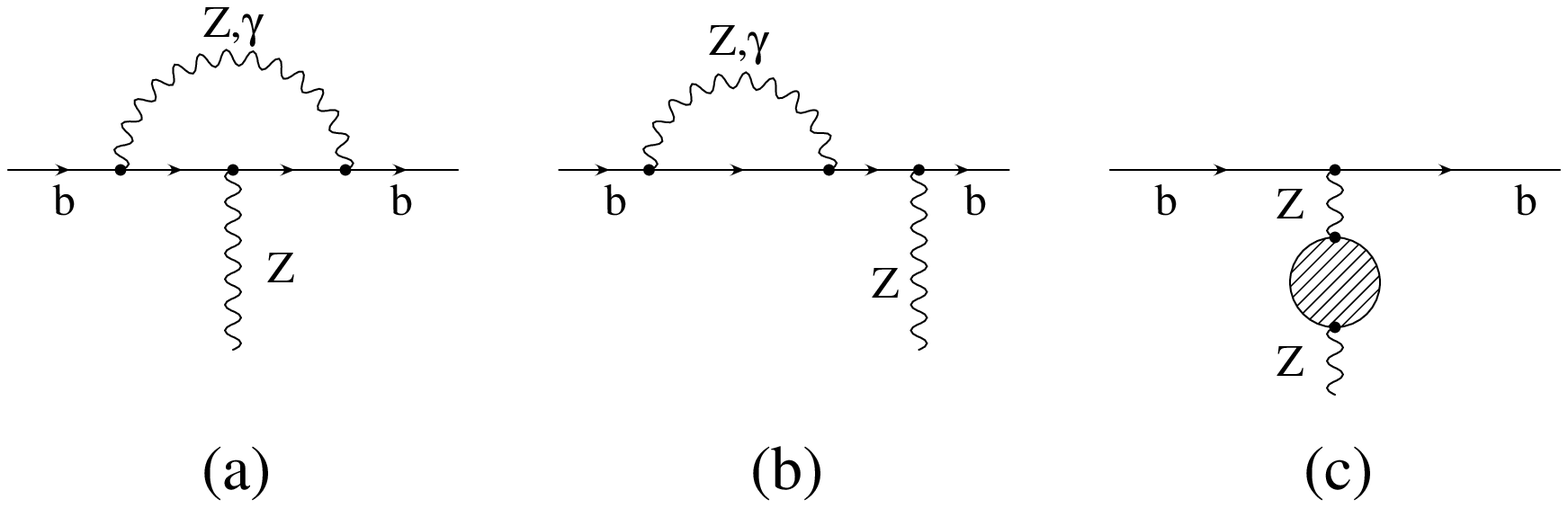}{Running of the vertex $Z$-$b$-$b$ in the effective
theory.}{zbb3}
\mafigura{16cm}{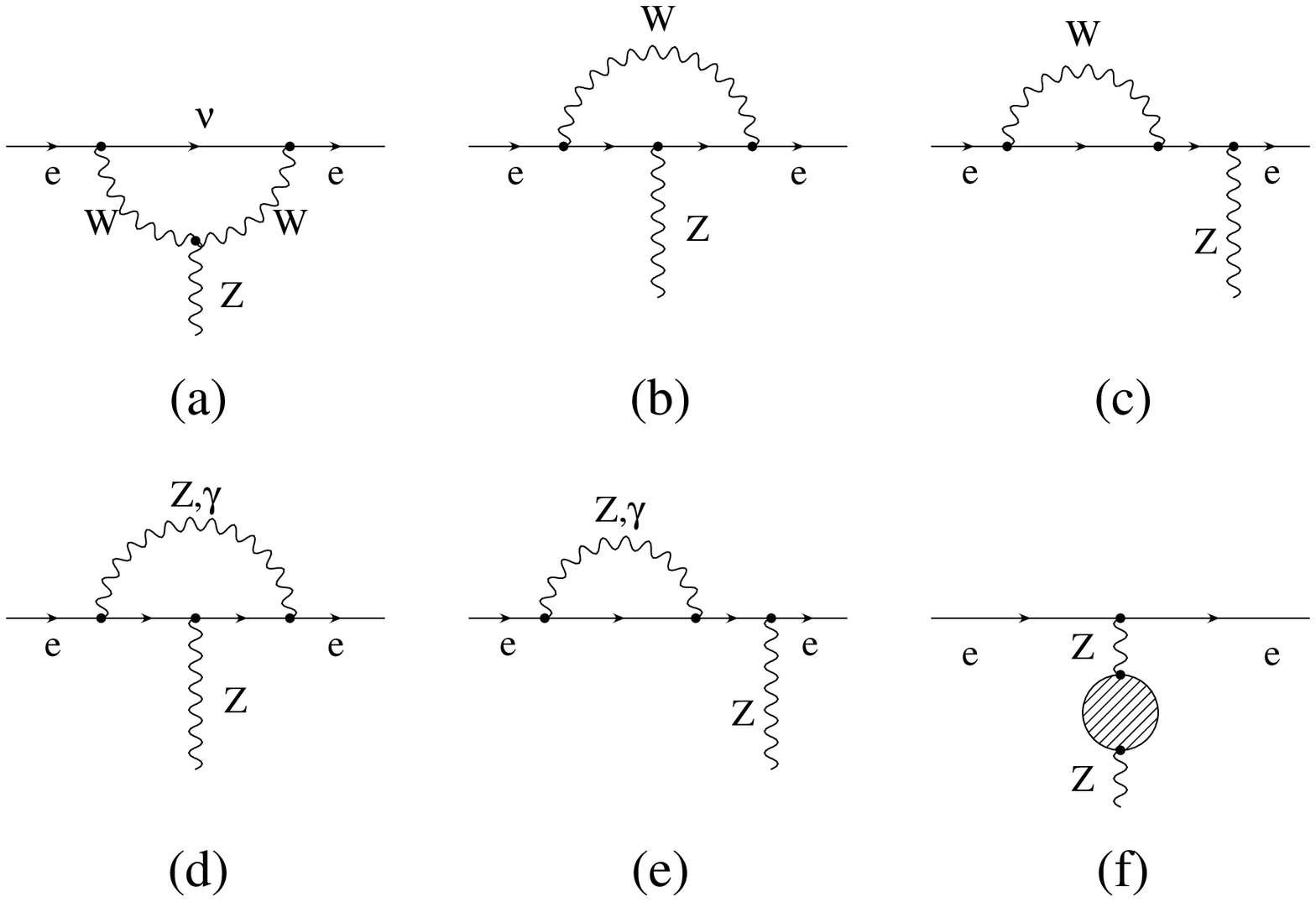}{Running of the vertex $Z$-$e$-$e$ in the effective
theory.}{zbb4}
\mafigura{13cm}{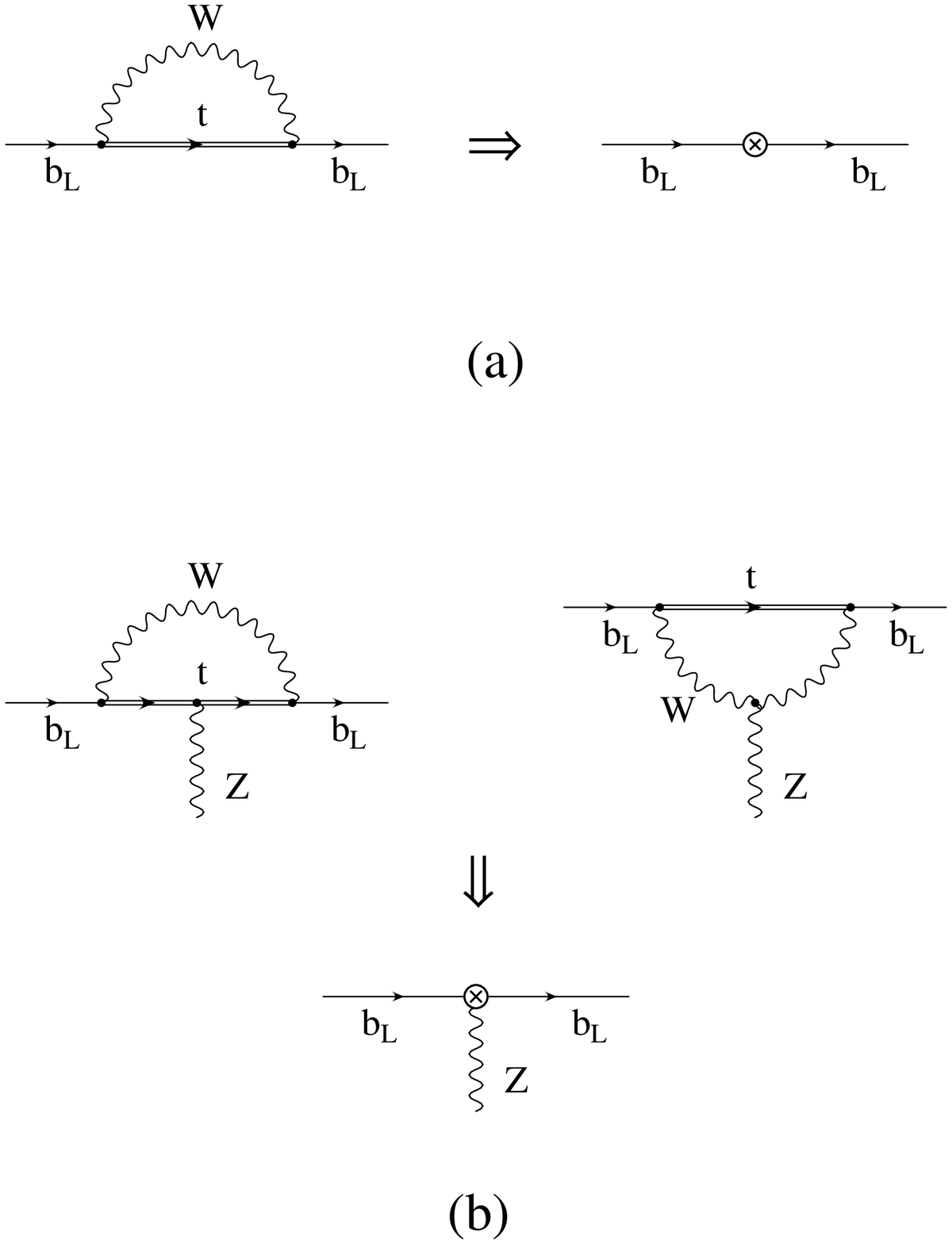}{One-loop matching: QCD switched off.}{zbb5}
\mafigura{13cm}{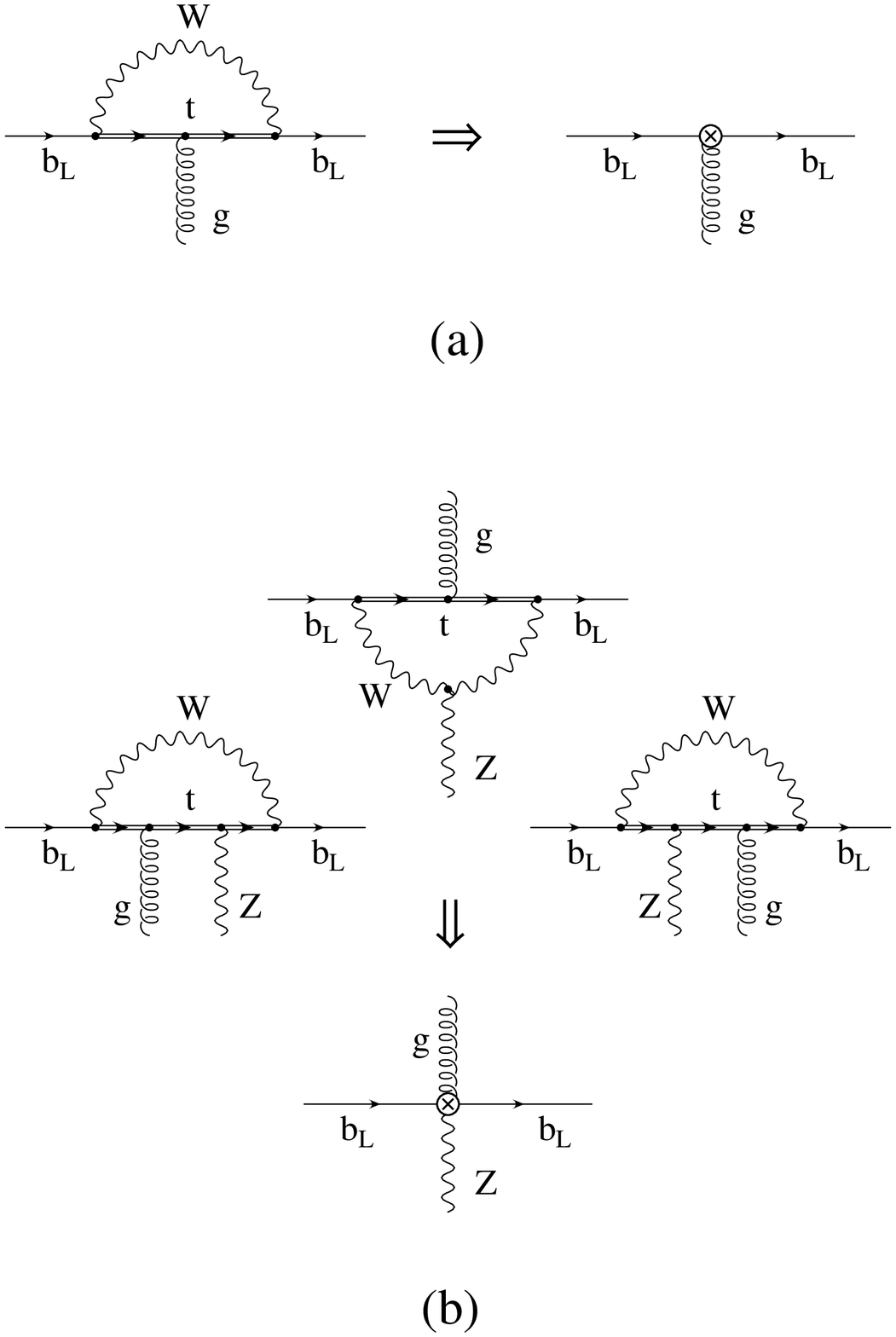}{One-loop matching: QCD switched on.}{zbb6}
\mafigura{12cm}{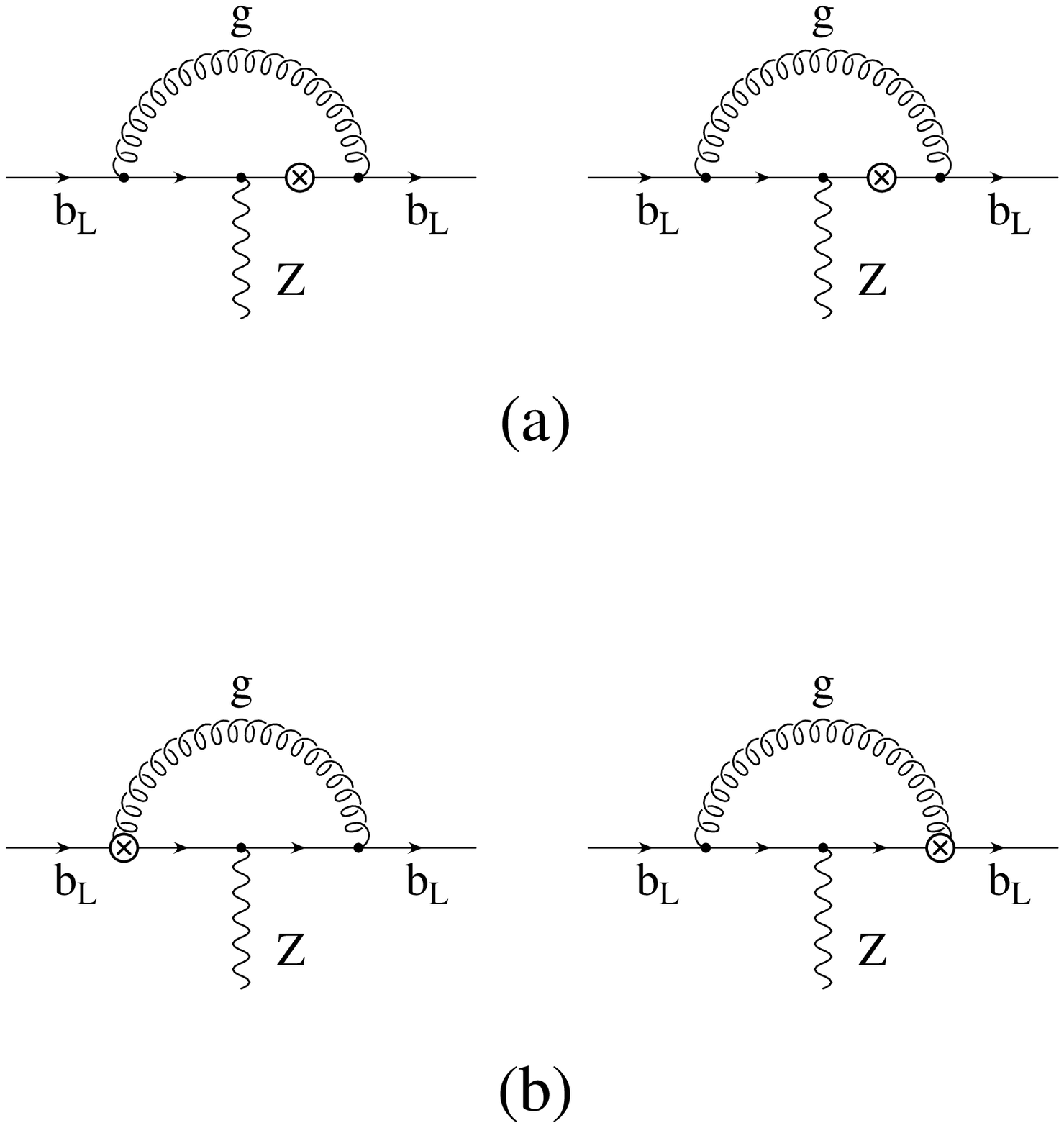}
{QCD running: Insertion of $b$-$b$ and $b$-$b$-$g$ operators.}{zbb7}
\mafigura{12cm}{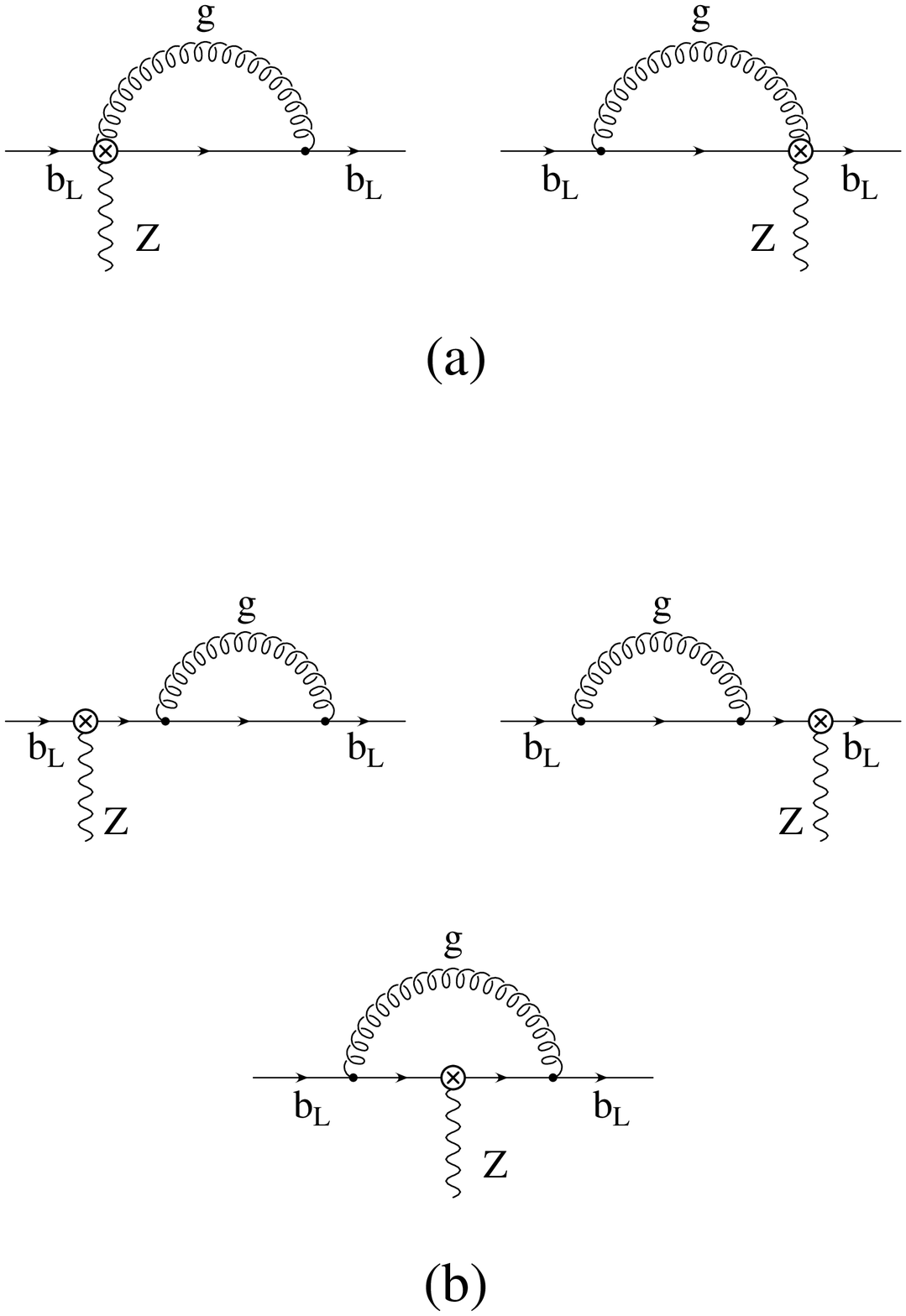}{QCD running: Insertion of $b$-$b$-$Z$ and
$b$-$b$-$Z$-$g$ operators.}{zbb8}
\mafigura{15cm}{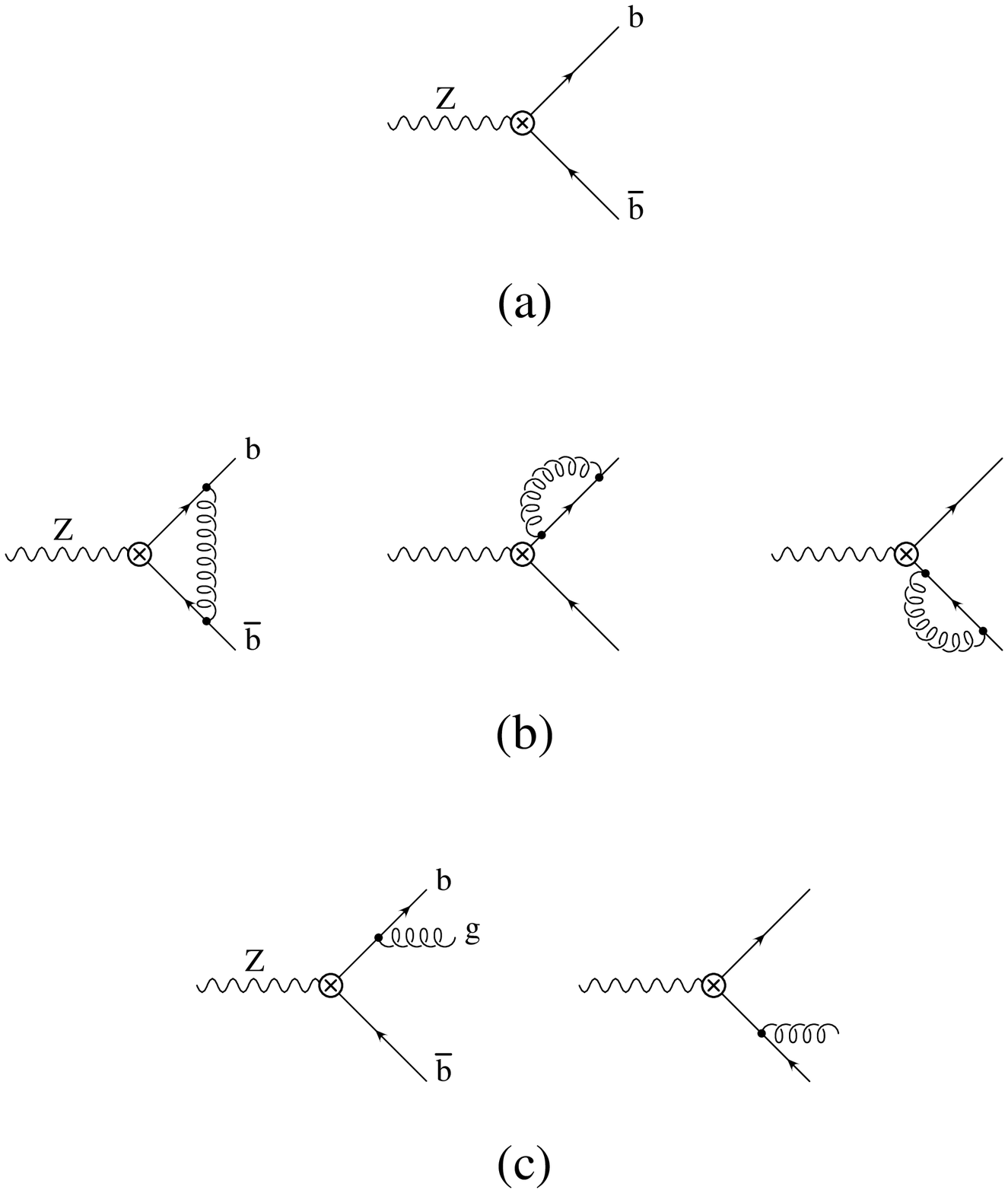}{Diagrams contributing to the matrix element
of $\zbb$ in the effective theory.}{zbb9}

\begin{thebibliography}{99}

\bibitem{marciano-sirlin}
W. Marciano and A. Sirlin, Phys. Rev. {\bf D22} (1980) 2695.

\bibitem{vanderBij-Veltman}
J.J. van der Bij and M. Veltman, Nucl. Phys. {\bf B231} (1984) 205.

\bibitem{Miquel}
See for instance, R.
Miquel, CERN-PPE/94-70. Talk given at the 22nd Symposium on ``Physics with
High Energy Colliders'', Tokyo, March 1994; and K. Abe et al., SLD
Collaboration, SLAC-PUB-6459, March 1994.

\bibitem{Kniehl-report}
See for instance, B.A. Kniehl, hep-ph-9410330. To appear in Int. J. Mod.
Phys. {\bf A}.

\bibitem{Tarasov-rho}
L. Avdeev,  J. Fleischer, S. Mikhailov and O. Tarasov, Phys. Lett. {\bf
B336} (1994) 560.

\bibitem{efts}
See, for instance, H. Georgi, Annu. Rev. Nucl. Part. Sci.
{\bf 43} (1993) 209; H. Georgi, "Weak Interactions and
Modern Particle Theory", The Benjamin/Cummings Pb. Co., 1984;
M. Bilenky and A. Santamaria, Nucl. Phys. {\bf B420} (1994) 47.

\bibitem{efts2}
In the context of a heavy top see for instance, H. Steger, E. Flores and
Y.-P. Yao, Phys. Rev. Lett. {\bf 59} (1987) 385; G.-L. Lin, H. Steger and
Y.-P. Yao, Phys. Rev. {\bf D44} (1991) 2139; ibid. Phys. Rev. {\bf D49}
(1994) 2414; F. Feruglio, A. Masiero and L. Maiani, Nucl. Phys. {\bf B387}
(1992) 523.

\bibitem{Hall} L. Hall, Nucl. Phys. {\bf B178} (1981) 75.

\bibitem{Witten} E. Witten, Nucl. Phys. {\bf B104} (1976) 445; {\bf B122}
(1977) 109.

\bibitem{Georgi-Cohen-Grinstein}
A. Cohen, H. Georgi and B. Grinstein, Nucl. Phys. {\bf B232} (1984) 61.

\bibitem{Grinstein-Wang}
B. Grinstein and M.-Y. Wang, Nucl. Phys. {\bf B377} (1992) 480.

\bibitem{Peris}
S. Peris, CERN-TH.7446/94. To appear in Phys. Lett. {\bf B}.

\bibitem{zbb-general}
A. Akhundov, D. Bardin and
T. Riemann, Nucl. Phys. {\bf B276} (1986) 1;
J. Bernab\'eu, A. Pich and A. Santamaria, Phys. Lett.
{\bf B200} (1988) 569; Nucl. Phys. {\bf B363} (1991) 326;
W. Beenakker and W. Hollik, Z. Phys. {\bf C40} (1988) 141.

\bibitem{RQCD}
See for instance, M. Consoli, W. Hollik and F. Jegerlehner, "Z Physics at LEP
1", CERN 89-08, edited by G. Altarelli, R. Kleiss and C. Verzegnassi, Sept.
1989.

\bibitem{Sirlin-Kniehl}
S. Fanchiotti, B. Kniehl and A. Sirlin, Phys. Rev. {\bf D48} (1993) 307.

\bibitem{Sirlin2}
A. Sirlin, BNL preprint, hep-ph 9403282, March 1994, and New York University
preprint NYU-TH-94/08/01.

\bibitem{Kwiatkowski-Steinhauser}
A. Kwiatkowski and M. Steinhauser, Kalsruhe preprint TTP94-14, Sept. 94.

\bibitem{Altarelli-Barbieri-Caravaglios-NPB405}
G. Altarelli, R. Barbieri and R. Caravaglios, Nucl. Phys. {\bf B405}
(1993) 3.

\bibitem{Fujikawa-Lee-Sanda-Pokorski-book11}
K. Fujikawa, B.W. Lee and A.I. Sanda, Phys. Rev. {\bf D6} (1972) 2923;
S. Pokorski, ``Gauge Field Theories'',  Cambridge University Press, 1987.

\bibitem{Arzt}
C. Arzt, Phys. Lett. {\bf B342} (1995) 189;
H. Simma, Z. Phys. {\bf C61} (1994) 67;
C. Grosse-Knetter, Phys. Rev. {\bf D49} (1994) 6709.

\bibitem{Martinelli-altres}
See for example, M. Ciuchini, E. Franco, G. Martinelli and L. Reina,
Nucl. Phys. {\bf B415} (1994) 403;
G. Cella, G. Curci, G. Ricciardi and A. Vicere, Phys. Lett. {\bf B248} (1990)
181; A.J. Buras, M. Jamin, P.H. Weisz, Nucl. Phys. {\bf B347} (1990) 491; P.
Cho and B. Grinstein, Nucl. Phys. {\bf B365} (1991) 279.

\bibitem{match-mt-2loops}
J. Fleischer, F. Jegerlehner, P. Raczka and O.V. Tarasov, Phys. Lett. {\bf
B293} (1992) 437; G. Degrassi, Nucl. Phys. {\bf B407} (1993) 271; G.
Buchalla and A.J. Buras, Nucl. Phys. {\bf B398} (1993) 285; K.G. Chetyrkin,
A. Kwiatkowski and M. Steinhauser, Mod. Phys. Lett. {\bf A7} (1993) 2785.

\bibitem{IR}
T. Kinoshita, J. Math. Phys. {\bf 3} (1962) 650;
T.D. Lee and M. Nauenberg, Phys. Rev. {\bf 133} (1964) 1549.

\bibitem{bile} For an explicit calculation including $b$-quark mass effects
see for instance,
M.~Bilenky, G.~Rodrigo and A.~Santamaria,
CERN-TH.7419/94, hep-ph/9410258, to appear in Nucl. Phys. {\bf B}.

\bibitem{djouadi-chetyrkin}
A. Djouadi, J.H. K\"uhn and P.M. Zerwas, Z. Phys. {\bf C46} (1990) 411;
K.G. Chetyrkin and J.H. K\"uhn, Phys. Lett. {\bf B248} (1990) 359.

\bibitem{Peris-unpublished}
S. Peris, unpublished.

\end{thebibliography}
\end{document}